\documentclass[cameraready]{Interspeech}
\usepackage{graphicx}
\usepackage{multicol}
\usepackage{multirow}

\title{Automatic Curation of Large-Scale, High-Quality, Multi-Category Music Source Separation Dataset}

\author[affiliation={1}, equalcontribution]{Yu}{Ji}
\author[affiliation={1}, equalcontribution]{Shuo}{Yang}
\author[affiliation={2}, equalcontribution]{Yuetonghui}{Xu}
\author[affiliation={1}]{Mengmei}{Liu}
\author[affiliation={1}]{Qiang}{Ji}
\author[affiliation={1}, correspondingauthor]{Zerui}{Han}

\address{
    $^1$ Xiaomi Inc., Beijing, China \\
    $^2$ Department of Music AI \& IT, Central Conservatory of Music, Beijing, China
}

\email{jiyu1@xiaomi.com, yangshuo7@xiaomi.com, 22a057@ccom.edu.cn, liumengmei@xiaomi.com, jiqiang@xiaomi.com, hanzerui@xiaomi.com}


\usepackage{comment}

\begin{document}

\maketitle

\begin{abstract}
The scale, audio quality, and granularity of instrument categories in datasets are critical factors limiting the performance of music source separation. Web crawling provides large-scale data but often introduces label noise and mismatches.
We propose an automated method for constructing large-scale, high-quality, multi-category music source separation datasets. The approach comprises 4 steps: (1) designing a fine-grained 7-category musical instrument taxonomy; (2) training instrument-specific single-source detectors to identify segments containing only the target instrument; (3) retrieving plenty of raw audio data from public platforms using multilingual keywords; and (4) automatically cleaning the raw data with the detectors to produce a high-quality dataset.
Our detector achieves 97.14\% average accuracy across seven categories. Training SOTA models with the cleaned data yields an average SDR improvement of 1.16,dB on standard benchmarks, confirming the dataset's effectiveness.\footnote{
    Codes and weights are available at: \url{https://github.com/scottishfold0621/ACMID.}}

\end{abstract}

\section{Introduction}

Music source separation (MSS) aims to isolate individual instrument stems from mixed audio signals\cite{defossez2019music, kong2021decoupling, rouard2022hybrid, luo2023music, lu2024music}. Recent advances in deep learning have significantly enhanced MSS performance\cite{rouard2022hybrid, luo2023music, lu2024music}. However, state-of-the-art supervised models, such as HT-Demucs \cite{rouard2022hybrid}, BS-Roformer\cite{lu2024music}, DTTNet\cite{chen2024music}, and SCNet\cite{10446651}, remain heavily dependent on large-scale, high-quality training datasets\cite{MUSDB18HQ, bittner2014medleydb, pereira2023moisesdbdatasetsourceseparation}. Current research faces two primary limitations: (1) manual dataset construction is labor-intensive and costly; (2) existing datasets are typically limited to coarse-grained 4-stem separation (vocals, bass, drums, and others), with insufficient data volume and lacking fine-grained instrument-level annotations.

To obtain more data, researchers turn to web-scale harvesting of audio from platforms such as YouTube, Bilibili\cite{miech2019howto100m, Chen_2023}, offering an enormous reservoir of candidate solo-instrument recordings.
Yet collecting single instrument tracks from such platforms inevitably introduces incorrect labels\cite{koo2023self, albert2022addressing, kang2023noise}: a query that originally intended to collect "guitar solo" may drag back a mix of bass, drums, and environmental noise, directly weakening the precise annotation on which current supervisory systems rely.

To overcome these challenges, particularly the scarcity of high-granularity data and label noise in web-crawled sources, we present an automated pipeline for constructing large-scale, high-quality, multi-category MSS datasets. Our main contributions are as follows:

\begin{itemize}
    \item We introduce an effective automated cleaning framework that achieves 97.14\% average accuracy in single-source detection, significantly mitigating label noise and audio-label mismatches in web-crawled data.
    \item We construct a large-scale, high-quality dataset with expanded 7-stem granularity focusing on musical instruments: ACMID(\textbf{A}utomatic \textbf{C}urated \textbf{M}usical \textbf{I}nstruments \textbf{D}ataset), addressing the longstanding shortage of fine-grained multi-instrument training data for MSS.
    \item Training state-of-the-art models with our cleaned dataset yields an average SDR improvement of 1.16 dB on standard benchmarks, demonstrating the superior quality and utility of the constructed data.
    \item We open-source the web crawling scripts, detector implementations, and pretrained weights to facilitate reproducibility and future research. 
\end{itemize}

\section{Methodology}
\label{sec:methodology}

\begin{figure*}[h]
    \centering
    \includegraphics[width=0.9\linewidth]{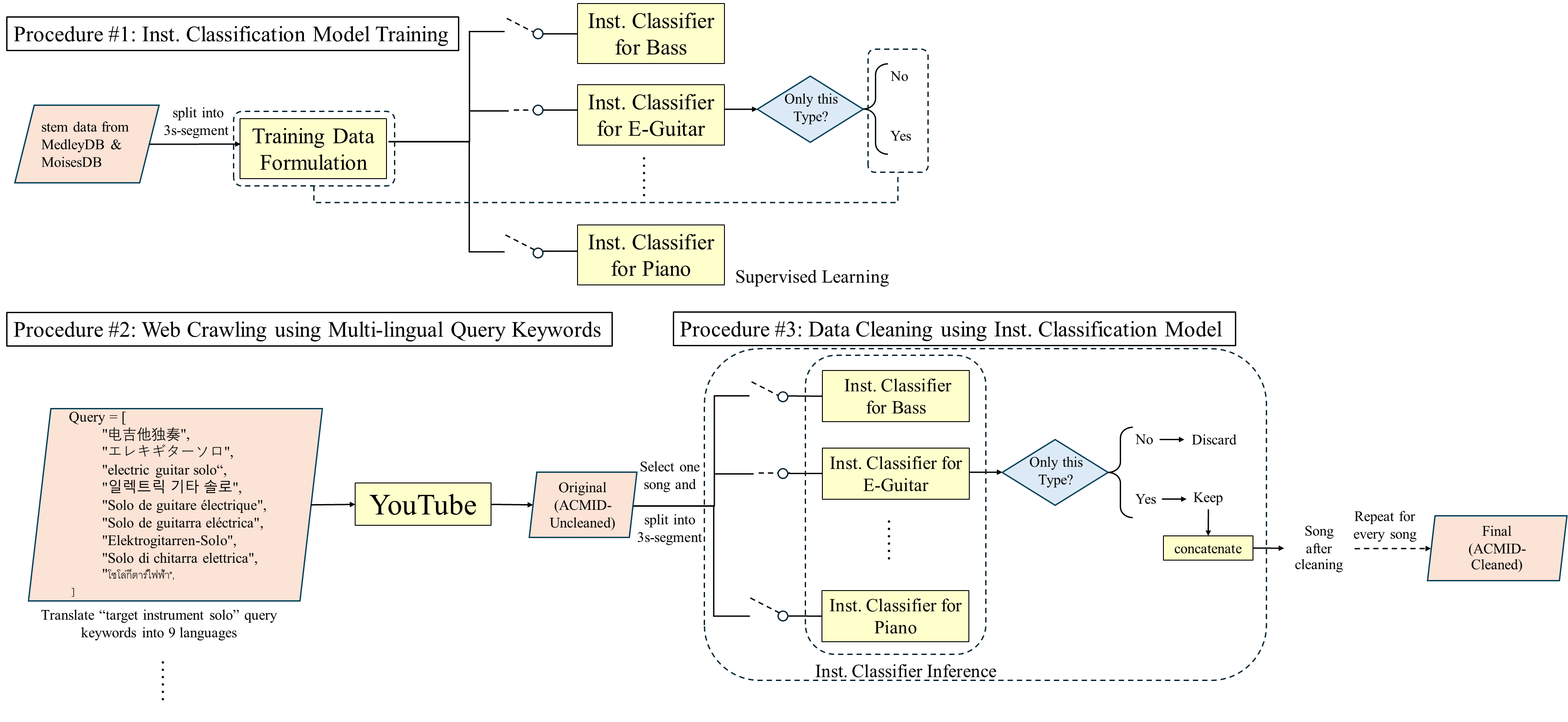}
    \caption{Our proposed procedure. }
    \label{fig:proposed_procedure}
\end{figure*}

In this section, we first define the musical instrument taxonomy adopted in our work. 
Then, we describe the method of our automatic curation of Large-Scale, High-Quality, Multi-Category music source separation dataset. It consists of 3 main procedures as shown in Fig. \ref{fig:proposed_procedure}:

Step 1: For each instrument, we train a binary classifier(with monophonic, 16khz input) based on a pre-trained audio encoder to detect whether an input segment contains only that instrument. These classifiers are used for subsequent automatic data cleaning.

Step 2: We collect raw videos of each instrument from Youtube using multilingual keyword queries. For all videos, we extract their audio tracks and convert them to stereo audio at a 48 kHz sample rate, forming the raw dataset (ACMID-Uncleaned).

Step 3: Each track in ACMID-Uncleaned is divided into 3-second segments. These segments are downsampled to monophonic 16 kHz and fed into the corresponding instrument-specific binary classifier, which identifies the time positions of pure target-instrument segments. Impure segments are discarded. To preserve high audio quality, the retained segments are extracted directly from the original 48 kHz audio at the detected time positions, concatenated with fade-in/fade-out transitions applied at boundaries to avoid artifacts, and saved as cleaned tracks. This automatic process is applied to all tracks, yielding the final high-quality dataset (ACMID-Cleaned).

\begin{table}[h]
\begin{tabular}{ll}
\hline
\hline
\textbf{Main Stem}   & \textbf{Sub Tracks}   \\ \hline \hline
Piano           &               \\ \hline
Drums           &               \\ \hline
Bass   &               \\ \hline
Acoustic Guitar &               \\ \hline
Electric Guitar &               \\ \hline
Strings         & Cello, Viola, Violin, Double Bass \\ \hline
Wind-Brass     & Trombone, Trumpet, Tuba, Euphonium, \\
                & French Horn, English Horn, Bassoon, \\
                & Clarinet, Contra Bassoon, Flute, \\ 
                & Oboe, Piccolo, Saxophone  \\ \hline
\end{tabular}
\caption{ACMID main-sub category taxonomy used to organize individual instrument categories into main categories. For Strings and Wind-Brass stem, we query every subcategory keywords. Others we only query the stem keywords.}
\end{table}

\subsection{Instrument Taxonomy}

Existing source separation datasets are broadly categorized into fixed-stem datasets (e.g., MUSDB18\cite{MUSDB18HQ}) and variable-stem datasets (e.g., MedleyDB\cite{bittner2014medleydb}). Fixed-stem datasets provide consistent source compositions across all mixtures, enabling standardized evaluation, fair cross-method comparisons, and simplified training procedures. However, most are confined to 4-stems (typically drums, bass, vocals, and others), resulting in inadequate separation granularity. Variable-stem datasets, while offering greater instrument diversity, often include overly fine-grained or highly specific stems that limit generalizability.

Given that the primary downstream applications of MSS are in pop music scenarios, we designed a fixed-stem dataset with seven common instruments: Piano, Drums, Bass, Acoustic Guitar, Electric Guitar, Strings, and Wind-Brass. Vocals were excluded due to their abundant representation in existing datasets.


\begin{figure}[htbp]
    \centering
    \includegraphics[width=0.5\linewidth]{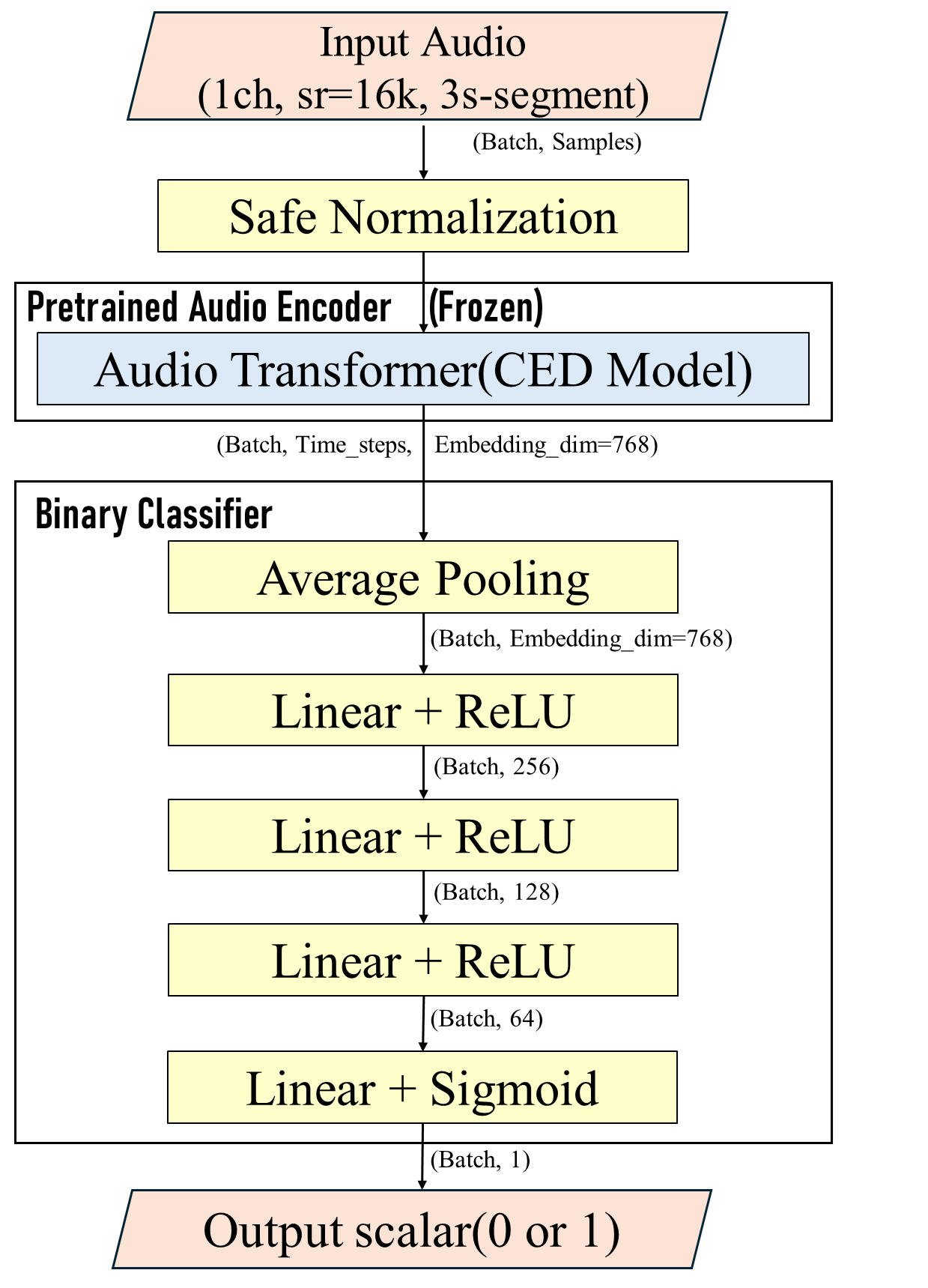}
    \caption{Our proposed audio classification model.}
    \label{fig:classification_modell}
\end{figure}

\subsection{Target Instrument Detector}\label{subsec:binary_classifier}

We train a target instrument detector for each instrument to determine whether the input segment contains only the target instrument. Each detector's architecture is shown in Fig. \ref{fig:classification_modell}. 
The input of the network is a single-channel 3s-segment with 16kHz sample rate. To avoid gradient instability caused by extreme amplitudes, safe normalization is performed on each signal segment:

$$x \leftarrow \frac{x}{\max\left(|x| \right) + \varepsilon},\quad \varepsilon = 10^{-9}$$

The normalized waveform is mapped to a high-level representation through a frozen pretrained audio encoder: Dasheng\cite{dinkel2024scalingmaskedaudioencoder}.
Dasheng is a self-supervised audio encoder based on masked autoencoder framework which competes well in music and environment classification.
The parameters of Dasheng are frozen during training and inference period of our binary classifier.

Based on the encoder's output, a binary classifier is designed to determine whether a segment is "clean" (label 1) or "impure" (label 0). 
Firstly, an average pooling is performed along the time dimension.
Then the feature dimension is reduced using three layers of "Linear $\to$ ReLU" with the hidden layer dimensions set to [D, D/2, D/4].
The final layer is a linear mapping that outputs scalar logits. After sigmoid processing, a probability estimate of the segment's purity is obtained. A threshold of 0.5 corresponds to a final 0 or 1 decision. 

\subsection{Web Crawling using Multilingual Query Keywords}\label{subsec:web_crawling}

We crawl publicly available content on YouTube and download relevant videos based on predefined query keywords.
To exclude noisy or irrelevant entries and retain only valid data specific to the target instrumental stems in the query results, we designed a standardized query template with the structure "instrument + solo". This template was further localized into nine languages to enhance cross-linguistic coverage, and the detailed translation of the query terms is provided in the bottom-left area of Fig. \ref{fig:proposed_procedure}.
For all videos, we keep their audio part, and transfer them into stereo audio with 48kHz sample rate.

\subsection{Data Cleaning using Instrument Classification Model}

After the initial filtering via web crawling with predefined multilingual query keywords in Sec.\ref{subsec:web_crawling}, we obtained the original ACMID-Uncleaned dataset. 
However, the data obtained via this crawling process still includes audio files containing non-target sounds.
Therefore, building upon the data obtained from crawling, we performed inference on the audio data of each instrument category using the corresponding pre-trained binary classification model, as shown in the bottom-right area of Fig. \ref{fig:proposed_procedure}. This step ensures that only pure segments of the target instrument are retained in the inferred results.

\begin{table*}[h]
\resizebox{\textwidth}{!}{
\begin{tabular}{c|ccccccc}
\hline
\multirow{2}{*}{\textbf{Model}} & \multicolumn{7}{c}{\begin{tabular}[c]{@{}c@{}}\textbf{Accuracy$\uparrow$} / \textbf{F1-Score$\uparrow$}\\ \textbf{Precision$\uparrow$} / \textbf{Recall$\uparrow$}\end{tabular}}                                                                                                                                                                                                                                                                                                                                                                                                                  \\ \cline{2-8} 
                       & \textit{Piano}                                                                 & \textit{Drums}                                                                 & \textit{Bass }                                                        & \textit{Acoustic Guitar}                                                       & \textit{Electric Guitar}                                                       & \textit{Strings}                                                               & \textit{Wind \& Brass}                                                         \\ \hline
Dasheng\cite{dinkel2024scalingmaskedaudioencoder} & \begin{tabular}[c]{@{}l@{}}\textbf{98.0\%}/\textbf{0.9791}\\ \textbf{0.9750}/\textbf{0.9832}\end{tabular} & \begin{tabular}[c]{@{}l@{}}\textbf{99.2\%}/\textbf{0.9919}\\ \textbf{1.0000}/0.9840\end{tabular} & \begin{tabular}[c]{@{}l@{}}\textbf{99.0\%}/\textbf{0.9897}\\ 0.9796/\textbf{1.0000}\end{tabular} & \begin{tabular}[c]{@{}l@{}}92.6\%/0.9234\\ \textbf{0.9911}/0.8643\end{tabular} & \begin{tabular}[c]{@{}l@{}}\textbf{96.0\%}/\textbf{0.9575}\\ \textbf{0.9585}/\textbf{0.9581}\end{tabular} & \begin{tabular}[c]{@{}l@{}}\textbf{97.8\%}/\textbf{0.9788}\\ \textbf{0.9732}/\textbf{0.9845}\end{tabular} & \begin{tabular}[c]{@{}l@{}}\textbf{97.4\%}/\textbf{0.9738}\\ \textbf{0.9959}/\textbf{0.9528}\end{tabular} \\ \hline

Music2latent\cite{pasini2024music2latentconsistencyautoencoderslatent} & \begin{tabular}[c]{@{}l@{}}95.8\%/0.9597\\ 0.9542/0.9653\end{tabular} & \begin{tabular}[c]{@{}l@{}}99.0\%/0.9899\\ 0.9919/\textbf{0.9880}\end{tabular} & \begin{tabular}[c]{@{}l@{}}98.6\%/0.9861\\ \textbf{0.9920}/ 0.9802\end{tabular} & \begin{tabular}[c]{@{}l@{}}\textbf{94.6\%}2/\textbf{0.9450}\\ 0.9355/\textbf{0.9545}\end{tabular} & \begin{tabular}[c]{@{}l@{}}94.2\%/0.9395\\ 0.9259/0.9534\end{tabular} & \begin{tabular}[c]{@{}l@{}}83.6\%/0.8225\\ 0.9634/0.7143\end{tabular} & \begin{tabular}[c]{@{}l@{}}90.8\%/0.9049\\ 0.9648/0.8521\end{tabular} \\ \hline

CNN\cite{11005342} & \begin{tabular}[c]{@{}l@{}}48.8\%/0.656\\ 0.488/1.000\end{tabular} & \begin{tabular}[c]{@{}l@{}}50.2\%/0.013\\ 0.012/0.015\end{tabular} & \begin{tabular}[c]{@{}l@{}}48.8\%/0.656\\ 0.488/1.000\end{tabular} & \begin{tabular}[c]{@{}l@{}}50.8\%/0.019\\ 0.018/0.021\end{tabular} & \begin{tabular}[c]{@{}l@{}}50.2\%/0.668\\ 0.502/1.000\end{tabular} & \begin{tabular}[c]{@{}l@{}}27.2\%/0.313\\ 0.302/0.326\end{tabular} & \begin{tabular}[c]{@{}l@{}}50.2\%/0.009\\ 0.008/0.010\end{tabular} \\ \hline

\end{tabular}}
\caption{Performance comparisons of binary classifiers for target instruments using different audio encoders}
\label{tab:encoder_ablation}
\end{table*}

\begin{table*}[h]
\centering
\small
\begin{tabular}{lccccccc}
\toprule
Dataset       & Acoustic Guitar & Bass   & Drums    & Electric Guitar   & Piano     & Strings     & Wind-Brass      \\
\midrule
MoisesDB\cite{pereira2023moisesdbdatasetsourceseparation} + MedleyDB\cite{bittner2014medleydb}      & 4.25    & 8.98   & 23.14   & 6.27     & 6.56     & 2.56     & 1        \\
Slakh\cite{manilow2019cuttingmusicsourceseparation}  & 73.68    & 95.22  & 88.93   & 159.43   & 145.85   & 163.39   & 101.94   \\
\textbf{ACMID-Uncleaned }       & 215.16   & 97.55  & 92.4    & 167.63   & 902.53   & 908.52   & 2259.72  \\
\textbf{ACMID-Cleaned}          & 15.57    & 11.6   & 16.04   & 10.45    & 483.21   & 97.84    & 102.64   \\
\textbf{ACMID-Cleaned(11h)}     & 10.45    & 10.45  & 10.45   & 10.45    & 10.45    & 10.45    & 10.45    \\
\bottomrule
\end{tabular}

\caption{Comparison of track durations between the obtained ACMID dataset and existing datasets (unit: hours)}
\label{tab:instrument_metrics}

\end{table*}

\begin{table*}[h]
\small
\begin{tabular}{l|cccccccc}
\hline
\multicolumn{1}{c|}{\multirow{2}{*}{Training Data}}                          & \multicolumn{8}{c}{SDR{[}dB{]}$\uparrow$}                                                                      \\ \cline{2-9} 
\multicolumn{1}{c|}{}                                                        & \textit{Piano} & Drums & Bass & Acoustic Guitar & Electric Guitar & Strings & Wind \& Brass & Average \\ \hline
MoisesDB\cite{pereira2023moisesdbdatasetsourceseparation} + MedleyDB\cite{bittner2014medleydb}                                                            & 4.36  & \textbf{8.06}  & 6.72         & 4.93            & 4.28            & 3.72    & 2.18          & 4.89    \\ \hline
Slakh\cite{manilow2019cuttingmusicsourceseparation}        & 2.36  & 5.88  & 3.88          & 1.61            & 0.52            & 0.70    & 1.29          & 2.32    \\ \hline
ACMID(Uncleaned)                                                             & 0.08  & 4.88  & 3.62          & 1.76            & 1.13            & 1.74    & 2.46          & 2.24    \\ \hline
ACMID(Cleaned)                                                               & 4.36  & 7.11  & 5.25          & 3.63            & 3.53            & 4.91    & 3.65          & 4.63    \\ \hline
ACMID(Cleaned-11-hours)                                                      & 3.15  & 5.77  & 3.51          & 3.06            & 2.04            & 3.29    & 2.05          & 3.41    \\ \hline
\begin{tabular}[c]{@{}l@{}}MoisesDB + MedleyDB\\ +ACMID(Cleaned) \end{tabular} & \textbf{6.07}  & 8.05  & \textbf{6.84}   & \textbf{5.49}    & \textbf{5.72}            & \textbf{5.93}    & \textbf{4.24}          & \textbf{6.05}    \\ \hline
\end{tabular}
\caption{MSS model results by instrument stems and datasets.}
\end{table*}

\section{Experiments and Results}
\label{sec:experiments}




We conduct two experiments. The first train and evaluate the target instrument detector.
The second apply the trained detector to clean the raw dataset(ACMID-Uncleaned), and the cleaned dataset(ACMID-Cleaned) is used to train and evaluate state-of-the-art (SOTA) source separation models, quantifying the performance.

\subsection{Training and Evaluation of Target Instrument Detector}\label{subsec:detector_training}

\subsubsection{Data Collection}

For training, we take the training part of MedleyDB\cite{bittner2014medleydb}(official split) and MoisesDB\cite{pereira2023moisesdbdatasetsourceseparation}(split into 8:1:1 with train/valid/test) and organize them into 7 categories. To ensure data balance between instruments, we add corresponding instrument parts of Bach10\cite{5404324}, ARME-Strings\cite{tomczak2023virtuoso}, SynthTab\cite{Zang_2024}, Maestro\cite{hawthorne2018enabling}
datasets into training data.
For validation and test, we use the valid/test part of MedleyDB and MoisesDB.

\subsubsection{Training Procedure}

The training dataset consists of balanced positive and negative samples, with approximately 10,000 segments each. Positive samples are obtained by randomly cropping 3-second segments from clean audio of the target instrument. Negative samples are generated by randomly mixing $k$ interference sources ($k \in [1, 5]$). For $k=1$, a single source is selected from the remaining six target instrument classes. For $k>1$, sources are randomly drawn with replacement from a candidate pool comprising other instruments, the target instrument itself, vocals, speech, and noise. Following mixing, all negative samples undergo LUFS normalization, with per-category target loudness ranges (e.g., piano: $-25$ to $-20$ LUFS) to discourage reliance on absolute loudness cues. 

Online data augmentation is applied to both positive and negative samples. Reverb is added with probability 0.5 using GPU-accelerated FFT convolution with one of 10 impulse responses (RT60 randomly sampled from 0.3 to 1.5\,s), simulating hall, room, and chamber acoustics. Dynamic compression is applied with probability 0.3, using thresholds of 0.1 to 0.3 and ratios of 2 to 4. EQ adjustments are performed with probability 0.5, involving randomized low-frequency boost, high-frequency attenuation, mid-frequency enhancement, or general parametric modifications. Finally, all samples are resampled to 16\,kHz for training.


\subsubsection{Ablation Study: Pretrained Audio Encoder Selection}

To select the optimal pretrained audio encoder for our proposed single-source instrument detector, we compared three candidates—Dasheng~\cite{dinkel2024scalingmaskedaudioencoder}, Music2Latent~\cite{pasini2024music2latentconsistencyautoencoderslatent}, and a CNN-based encoder~\cite{11005342}—within the same binary classification framework. Performance was evaluated across the seven instrument categories using Accuracy, Precision, Recall, and F1-Score.

Results are presented in Table~\ref{tab:encoder_ablation}. Dasheng consistently achieves the highest scores across nearly all categories and metrics, followed by Music2Latent with competitive performance on select instruments, while the CNN-based encoder ranks lowest overall.

We therefore integrate Dasheng as the audio encoder in our final detector. The proposed single-source instrument detector attains an average accuracy of 97.14\% across the 7 instrument classes.

\subsubsection{Ablation Study: Classifier's Threshold Selection}

\begin{figure}[!h]
    \centering
    \includegraphics[width=0.95\linewidth]{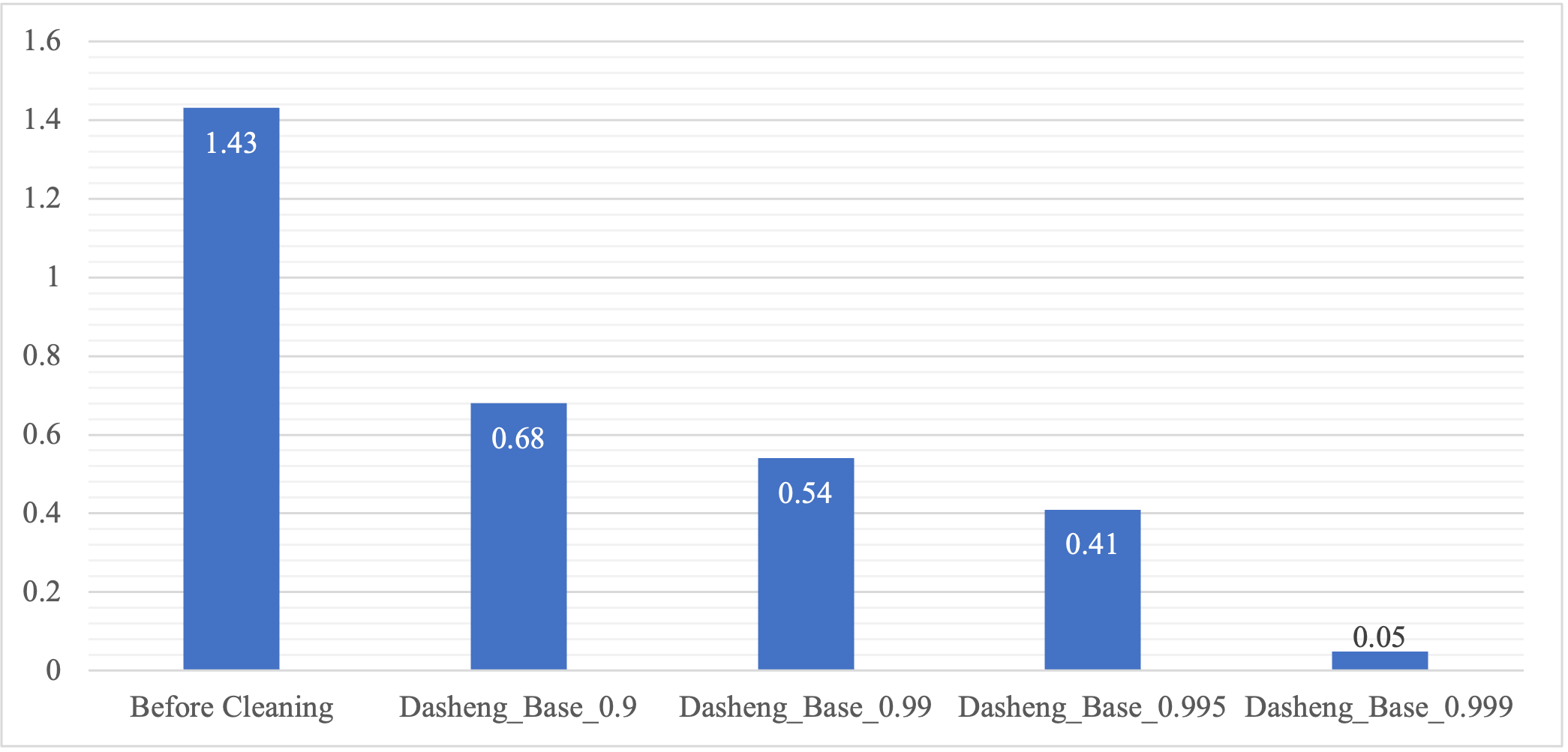}
    \caption{Duration Distributions Before and After Cleaning Under Different Thresholds(unit: hours)}
    \label{fig:duration_after_cleaning}
\end{figure}

As shown in Fig. \ref{fig:classification_modell}, the target instrument detector uses a sigmoid output compared against a threshold to determine whether a segment contains purely the target instrument. This threshold governs the quality-quantity trade-off in the cleaned data. To find the optimal balance, we tested thresholds of 0.9, 0.99, 0.995, and 0.999 with the Dasheng-Base\cite{dinkel2024scalingmaskedaudioencoder} encoder on 10 crawled songs covering all the 7 target instruments. Duration distributions before and after cleaning are shown in Fig. \ref{fig:duration_after_cleaning}.

To evaluate cleaning reliability, three graduate students majoring in sound recording are paid to conduct subjective evaluation. The 10 songs are 1.43 hours before cleaning. Post-cleaning results are as follows:
(a) threshold 0.9: 0.68 hours retained, with minor non-target noise;
(b) threshold 0.99: 0.54 hours retained, with markedly reduced noise;
(c) threshold 0.995: 0.41 hours retained, with minimal interference in few segments;
(d) threshold 0.999: 0.05 hours retained, with virtually no interference.

Although 0.999 yielded the cleanest data, the retained duration (0.05 hours) was insufficient for subsequent MSS model training. Thus, we adopted 0.995 as the final threshold.

\subsection{Impact of ACMID on Music Source Separation Systems}

To evaluate the impact of our curated dataset on multitrack source separation (MSS), we trained state-of-the-art models on six different configurations: (1) MoisesDB + MedleyDB; (2) Slakh~\cite{manilow2019cuttingmusicsourceseparation}, a large-scale synthetic dataset; (3) ACMID-Uncleaned (raw crawled data); (4) ACMID-Cleaned (fully cleaned version); (5) ACMID-Cleaned-11h, a balanced subset with approximately 10.45 hours per instrument category; and (6) ACMID-Cleaned combined with MoisesDB + MedleyDB. Dataset durations are reported in Table~\ref{tab:instrument_metrics}.

Models were validated and tested on the validation and test splits of MoisesDB and MedleyDB, using the same protocol as described in Section~\ref{subsec:binary_classifier}. Performance was quantified using the signal-to-distortion ratio (SDR).

Results are presented in Table~3. As anticipated, training solely on MoisesDB + MedleyDB yielded the strongest standalone performance, benefiting from domain alignment with the test set. In contrast, Slakh performed poorly despite its substantial duration, underscoring the limitations of synthetic data with distribution mismatch in real-world scenarios.

Within our datasets, ACMID-Uncleaned exhibited low performance due to prevalent label noise and contamination. ACMID-Cleaned, however, demonstrated substantial improvements across all instruments, validating the effectiveness of the proposed automated cleaning pipeline. The balanced ACMID-Cleaned-11h subset, designed to mitigate minor category imbalances, slightly underperformed the full ACMID-Cleaned, suggesting that increased data volume outweighs perfect balance in this setting. Finally, augmenting the baseline MoisesDB + MedleyDB with ACMID-Cleaned achieved the highest overall results, delivering an average SDR gain of 1.16,dB and highlighting the complementary value of our high-quality, real-world data.


\section{Conclusions}
\label{sec:conclusions}



In this work, we propose ACMID, a large-scale, high-quality, multi-instrument dataset featuring fixed 7-stem separations, constructed through an automated curation pipeline. 
The proposed single-source instrument detector for automatic cleaning achieves an average accuracy of 97.5\% across the seven instrument categories. 
Experiments demonstrate that training state-of-the-art source separation models on the cleaned ACMID data yields consistent performance improvements, with an average SDR gain of 1.16\,dB over models trained on unfiltered data.

To ensure reproducibility, we openly release the data crawling scripts, the instrument detection model code, and pretrained weights. Moreover, the proposed method is generalizable and can be readily extended to additional instrument categories, including traditional Chinese folk instruments.

\section{Use of generative AI tools}

 Generative AI tools have been used for editing and polishing, not for writing any major part.

\bibliographystyle{IEEEtran}
\bibliography{mybib}

\end{document}